**Title:** Validity of Complete Case Analysis Depends on the Target Population

**Authors:** Michael Webster-Clark[1,2] and Rachael K. Ross[3]

**Affiliations:**

1. Department of Epidemiology and Biostatistics, McGill University, Montreal, QC

2. Department of Epidemiology, University of North Carolina at Chapel Hill, Chapel Hill, NC

3. Department of Epidemiology, Columbia University, New York, NY.

**Abstract:**

Missing data is a pernicious problem in epidemiologic research. Research on the validity of complete case analysis for missing data has typically focused on estimating the average treatment effect (ATE) in the whole population. However, other target populations like the treated (ATT) or external targets can be of substantive interest. In such cases, whether missing covariate data occurs within or outside the target population may impact the validity of complete case analysis. We sought to assess bias in complete case analysis when covariate data is missing outside the target (e.g., missing covariate data among the untreated when estimating the ATT). We simulated a study of the effect of a binary treatment X on a binary outcome Y in the presence of 3 confounders C1-C3 that modified the risk difference (RD). We induced missingness in C1 only among the untreated under 4 scenarios: completely randomly (similar to MCAR); randomly based on C2 and C3 (similar to MAR); randomly based on C1 (similar to MNAR); or randomly based on Y (similar to MAR). We estimated the ATE and ATT using weighting and averaged results across the replicates. We conducted a parallel simulation transporting trial results to a target population in the presence of missing covariate data in the trial. In the complete case analysis, estimated ATE was unbiased only when C1 was MCAR among the untreated. The estimated ATT, on the other hand, was unbiased in all scenarios except when Y caused missingness. The parallel simulation of generalizing and transporting trial results saw similar bias patterns. If missing covariate data is only present outside the target population, complete case analysis is unbiased except when missingness is associated with the outcome.

Missing data is a major threat to the validity of both experimental and observational studies. Missing data is often categorized as missing completely at random (MCAR), at random with respect to measured variables (MAR), or missing not at random (MNAR).(1) The simplest and most common method for handling missing data is complete case analysis (CCA) or excluding individuals with missing data from the analysis. Validity of CCA does not conform to the classification MCAR, MAR, and MNAR; rather validity of CCA depends on the underlying causal structure, the parameter of interest, and effect measure modification.(2-5)

Work examining CCA validity has focused on estimating conditional associations or the average treatment effect (ATE) where the entire population is the target (i.e., when everyone in the analytic population is part of the target population and vice versa). However, other estimands are frequently of substantive interest.(6) For example, we may want to estimate the average treatment effect in the treated (ATT), estimate the outcome risk under standard of care for single-arm trial participants from a real-world comparator,(7) or transport treatment effect estimates from a randomized trial to an external target.(8, 9) For these other estimands, the analytic and target populations are only partially overlapping. It is not clear the extent to which estimator properties such as CCA validity may be different for these other estimands.

We illustrate here that CCA is generally valid when missing baseline covariate data is restricted to the analytic population that is not part of the target population (i.e., the covariate data for the target population are fully observed), except when the outcome causes or is associated with missingness independently of the partially measured covariates. First, we demonstrate this in the context of estimating the ATT in which there is covariate missingness in the untreated but not the treated (target population) by contrasting observed bias in the ATT with bias in the ATE. Next, we show that a similar principle applies if data on effect measure modifiers are missing for some trial patients when transporting estimates to an external population.

**METHODS**

Definitions

*Study population:* A study population is the group of people that researchers have sampled meeting eligibility criteria for the study.

*Target population:* A target population is a group of people whom researchers wish to estimate the parameter of interest in and it can be internal or external to the study population.(10, 11) Some common target populations include the full study population, treated individuals in the study population, those who were eligible for a clinical trial, or all real-world users of a medical intervention.

*Analytic population:* The analytic population is the group of individuals used to estimate the parameter of interest in the target population. Everyone in a target population is generally a member of the analytic population, but not all members of the analytic population are part of a given target population. For example, when estimating the ATT, the untreated are members of the analytic, but not target, population. Similarly, when transporting a treatment effect to an external target, the study population is part of the analytic, but not target, population.

Average treated effect in the treated and average treatment effect

Suppose we are interested in estimating the ATT, but some untreated individuals are missing data on a confounding variable $M$. Let us consider four different potential mechanisms for untreated individuals to be missing $M$: **Type 1**, where the probability of missing M is completely random with respect to the values of other confounders, the value of $M$, and the outcome; **Type 2**, where the probability of missing $M$ varies based on the values of other confounders; **Type 3**, where the probability of missing $M$ varies based on its own value; and **Type 4**, where the probability of missing $M$ varies based on the value of the outcome (e.g., outcome and missing $M$ share a common cause, or the outcome causes someone to be missing $M$). These four types correspond roughly to MCAR, MAR (with respect to other covariates), MNAR, and MAR (with respect to the outcome), respectively, within strata of the treatment. **Figure 1** presents directed acyclic graphs (DAGs) corresponding to each type for a hypothetical variable $M$ in the presence of two other confounding variables $N$ and $O$.

*Simulation design:* We conducted a simulation study to assess bias in the complete case analysis when estimating the ATT and ATE when only untreated individuals are missing covariate data. We simulated 2,000 replicates of 100,000 patients treated with a binary treatment $X$ followed for a binary outcome $Y$ in the presence of three binary confounding variables $M$, $N$, and $O$ under each of the four DAGs in **Figure 1**, where only $M$ had the potential to be missing and that missingness occurred only in those with $X = 0$, the untreated (full parameters included in **Table 1**). First, we determined the true ATT and ATE risk differences in the full simulated data (2 billion patients) using weighting (odds weights for the ATT, inverse probability weights for the ATE) before inducing missingness in $M$ using data. Next, for each scenario and within each replicate, we estimated the ATT and ATE risk difference among complete cases using odds weights (for the ATT) and inverse probability of treatment weights (for the ATE). We then compared the mean and standard errors of these estimates across all 2,000 replicates to the truth.

Transporting and generalizing estimates

Suppose we are interested in estimating the treatment effect in a target population that is, at least partially, external to a randomized trial population, but some individuals in the trial population are missing data on an effect measure modifier $M$.

Hereafter, we refer to the average treatment effect in the external target population as the AT-TRANS (as in TRANSported).(12) The average treatment effect in a hypothetical target population that includes the study population is the AT-GEN (as in GENeralized).(13)

Let us again consider four different potential mechanisms (that mirror those in the previous section) for individuals in the trial to be missing $M$: **Type 1**, where they are missing data on $M$ completely randomly; **Type 2**, where the probability of missing $M$ varies based on the values of other variables; **Type 3**, where the probability of missing $M$ varies based on the value of $M$; and **Type 4**, where the probability of missing $M$ varies based on the value of the outcome (either because of a common cause of missing $M$ and the outcome or an effect of the outcome on missing $M$). **Supplemental Figure 1** includes four DAGs for the hypothetical trial populations.

*Simulation design:* We conducted a second simulation study to assess bias in the complete case analysis when estimating the AT-TRANS and AT-GEN when missing covariate data is restricted to the trial participants. We simulated 2,000 replicates of 100,000 trial participants and 100,000 target population members. The trial participants were randomly treated with a binary exposure $X$ and followed for a binary outcome $Y$ in the presence of three binary EMMs $M$, $N$, and $O$ under each of the four DAGs in **Supplemental Figure 1,** where only $M$ had the potential to be missing (full parameters in **Supplemental Table 1**). $M$ was measured in all non-trial participants.

To obtain the true AT-TRANS and AT-GEN, we used inverse odds of sampling weights (for the AT-TRANS) and inverse probability of sampling weights (for the AT-GEN) based on $M$, $N$, and $O$ before missing data was induced and using all simulated data (n=2 billion). Next, for each scenario and within each replicate, we estimated the AT-TRANS and AT-GEN risk difference among complete cases with measured values of $M$ using inverse odds weights (for the AT-TRANS) and inverse probability weights (for the AT-GEN). We then compared the mean and standard errors of these estimates across all 2,000 replicates to the truth.

**RESULTS**

**ATT and ATE**

*Simulation:* **Table 2** presents the estimated ATT and ATE risk differences and their standard deviations under each type of missingness. For the ATT, CCA was unbiased under **Type 1**, **Type 2**, and **Type 3** missingness but was biased under **Type 4** missingness. As expected, CCA was less precise than if all data were observed. In contrast, CCA was biased for the ATE under all four types of missingness.

*Theoretical explanation:* The ATT is a contrast of the observed outcomes (under treatment) for the treated population and the counterfactual outcomes (under no treatment) for the treated population. Untreated individuals are used to estimate counterfactual outcomes for treated individuals (the target) based on the covariate distribution in the treated; as long as the covariate distribution is preserved in the treated, and the covariate-outcome associations are preserved in the untreated, the complete case analysis is unbiased. Using our definitions above, the untreated are part of the **analytic**, but not the **target**, population.

If $M$'s missingness is **Type 1** (i.e., completely random in the untreated), CCA will reduce the size of the untreated population used to estimate counterfactual outcomes in the treated individuals however the covariate-outcome associations in the untreated and the covariate distribution in the treated are preserved. There may be a loss of precision, but there is no expectation of bias.

When $M$'s missingness is **Type 2, or Type 3,** although the covariate distribution in the complete cases of the untreated differs from the full untreated population, this difference has no impact on whether the untreated are able to provide accurate estimates of the counterfactual outcomes in the treated after adjusting for $M$, $N$, and $O$. This difference is only problematic when the untreated are part of the target population, and they are not part of the target population for the ATT.

**Figure 2** provides a graphical explanation for why the complete case analysis ATT is not biased when missingness is **Type 2** or **Type 3**. If we have fully measured covariate data in the treated, our "target" covariate distribution is constant regardless of any missingness in the untreated; complete case analysis may change the weights each individual receives, but **not** the overall probability of the outcome after weighting. **Supplemental Proof 1** includes a proof of these findings.

The reason **Type 4** missingness of $M$ generates bias is because the covariate-outcome associations are biased. As a result of the open path from the outcome to missing $M$, standardizing or building models with the complete cases among the untreated will lead to an inaccurate estimate of the counterfactual outcomes within the treated.

Unlike for the ATT, untreated individuals are part of the target population for the ATE. Thus, CCA may be biased when the covariate distribution of the full study population is altered when we restrict to complete cases. This is even the case when missingness is **Type 1** (i.e., independent of the outcome and any confounding covariates except via exposure). In such cases, CCA excludes a random portion of the untreated from the analysis. Normally, the ATE risk difference is a weighted average of the risk differences among the treated and untreated with weights corresponding to the prevalence of treatment in the study population. By excluding a random portion of the untreated with missingness, the weights are different and can generate bias in the ATE. That said, there will not be bias if the treatment effect is the same in the treated and untreated (i.e., there are no effect measure modifiers associated with treatment).

AT-TRANS and AT-GEN

*Simulation:* **Table 3** presents the estimated AT-TRANS and AT-GEN risk differences and their standard deviations under each type of missingness. Results paralleled those for the ATT and ATE: CCA was biased for the AT-GEN under all types of missingness while CCA was biased for the AT-TRANS only under Type 4 missingness. Again, CCA was less precise than if all data were observed.

*Theoretical explanation:* When estimating the AT-TRANS, the trial participants are part of the analytic population but not the target population (analogous to the ATT); for the AT-GEN, the trial participants are part of both the analytic and target populations (analogous to the ATE). While **Type 2** and **Type 3** missingness mean that the complete cases among the trial participants do not reflect the true distribution of $M$ among all trial participants, estimating the AT-TRANS means that trial participants are reweighted to have the same distribution of $M$ as the external population regardless. On the other hand, all four types of missingness bias the AT-GEN specifically because correctly estimating the true distribution of $M$ in the trial participants is essential for reweighting the trial participants to the correct distribution of $M$ in the combined trial and target populations.

**Table 4** summarizes which types of missingness threaten the validity of complete case analysis for the ATT, ATE, AT-TRANS, and AT-GEN.

**DISCUSSION**

CCA can be an unbiased approach when missing data are restricted to individuals who are part of the analytic population but not part of the final target population. This is true whether the covariates are variables that need to controlled for eliminate confounding or EMMs that need to be adjusted for to estimate an externally valid treatment effect. CCA is not a universal solution; if some individuals are missing data because of the outcome, or there is a common cause of missing the covariate and the outcome, CCA may be biased. In such cases, other approaches such as weighting or multiple imputation may be unbiased and preferred.

*Relevance to studies with external comparators:* In epidemiologic studies embedded in single data sources like longitudinal surveys, insurance claims, registries, or medical records, missing covariate data is not likely to be restricted to one treatment group. In contrast, missingness restricted to one treatment group may occur in studies using external comparator arms. In such studies (primarily conducted when studying medical interventions for rare conditions), researchers combine data from a phase II study in which all participants received the investigative treatment in a highly controlled environment with external real-world data from individuals who received the standard of care or a non-investigative treatment.(7) Because these are non-experimental studies, the investigators generally must adjust for systematic differences between the highly selected study participants and the real-world data sample to avoid confounding. This is routinely done via propensity score matching to yield data that closely resembles a block randomized trial, meaning the investigators are generally estimating the treatment effect in the phase II study participants, or "what if the trial participants had, counter to fact, been treated with the treatment received by the real-world patients?"

The high level of regulatory scrutiny for many of these phase II studies means that data on covariates for the phase II study participants are collected and documented rigorously.(14) On the other hand, collection and documentation of the real-world data for the comparator group is often less rigorous. Thus, the comparator group may be missing some covariate data (i.e., the analytic population that are not part of the target population). We have illustrated that in such cases CCA is unbiased as long as the missingness remains conditionally independent of the outcome. Moreover, the size of the comparator group in these types of studies is often much larger than the phase II study such that restricting to complete cases may have negligible impact on precision.

*Relevance to transporting effects from randomized trials:* Understanding the AT-TRANS is essential to evaluating the potential public health impact of randomized trials. Because we cannot go back in time and change the treatment that trial participants received, including them in the target population (as we do in the AT-GEN) has limited practical implications. Moreover, data collection in trials that are big enough to meaningfully estimate an AT-TRANS (i.e., phase III clinical trials) is often focused on the primary outcome, adherence, and side effects; in contrast collection of EMMs may be less rigorous. It is possible that a substantial portion of randomized trial participants will be missing data on some EMMs. If we determine that the EMMs are important and prospectively gather target population data, we can better capture them in the target population and achieve negligible missingness; indeed, some data sources (like insurance claims databases) generally capture comorbidities with 0% missingness (albeit some misclassification), since they are defined based on the presence or absence of a diagnosis code.

*Limitations:* Our simulations were simple and focused on demonstrating the presence or absence of bias, not its general magnitude or the specific relationships between each simulation parameter and the final bias we observed in each analysis. We provide proofs that support bias illustrated in our simulations. Additionally, in our simulation examining the ATT and ATE, the confounding variables modified the treatment effect on the additive scale. If none of the confounding variables were modifiers, then CCA is unbiased for the ATE.

We also focused on four specific target populations: the ATT, the ATE, the AT-TRANS, and the AT-GEN. Other target populations may be of interest to researchers, including specific subgroups, the population targeted by specific types of weights like overlap weights or balancing weights, or the local average treatment effect estimated by instrumental variable analyses. While some of these populations have clearly defined analytic and target populations, others do not, making it difficult to identify whether or how our conclusions would extend to researchers estimating effects in these alternative target populations.

*Conclusion:*

Complete case analysis can be a valid approach to missing covariate data if missingness is limited to members of the analytic population who are not included in the target population. Casually extending conclusions about bias from situations where the analytic and target populations completely overlap, like the ATE, can be problematic.

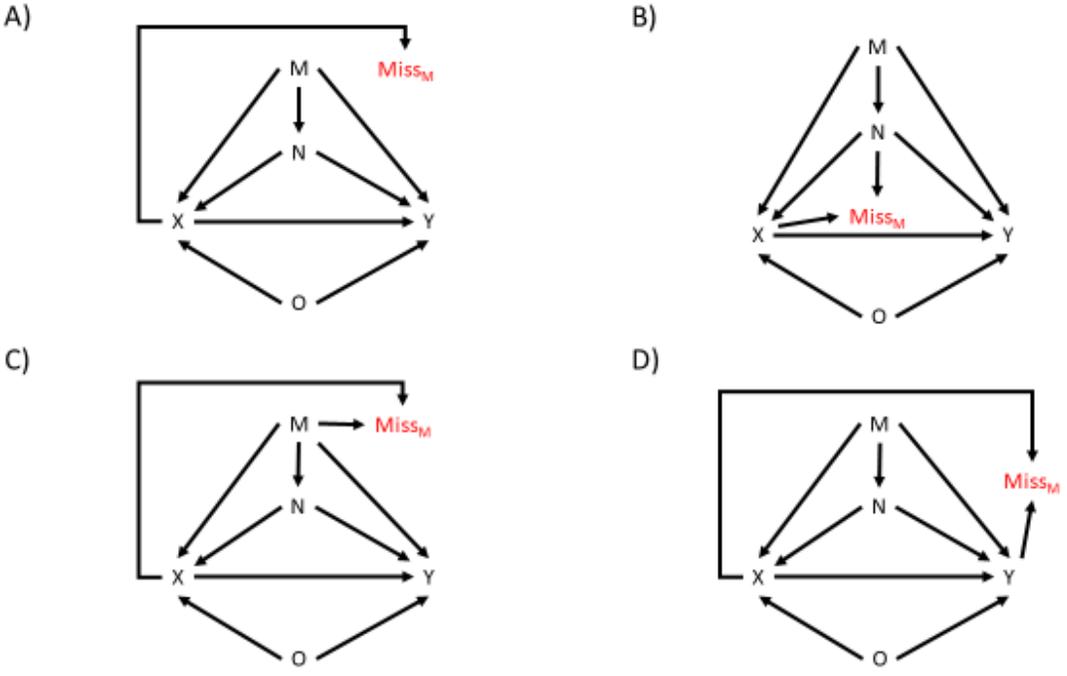

**Figure 1 Legend:** Four directed acyclic graphs corresponding to four potential types of missingness of a covariate *M* that is missing only within the untreated. Type 1 corresponds to panel A, Type 2 to panel B, Type 3 to panel C, and Type 4 to panel D.

**Table 1:** Simulation parameters for the simple ATT simulation

| Parameter | Value or value(s) |
|---|---|
| Number of individuals per simulation replicate | 100,000 |
| Number of replicates | 2,000 |
| P(C1) | 0.5 |
| P(C2 \| C1=1) | 0.3 |
| P(C2 \| C0=0) | 0.7 |
| P(C3) | 0.2 |
| P(X=1) | $0.1 + 0.2*C1 + 0.2*C2 + 0.2*C3$ |
| P(Y=1) | $0.05 + 0.1*C1 + 0.1*C2 + 0.1*C3 + 0.1*C1*C2 + 0.15*X*C1 + 0.15*X*C2 + 0.10*X*C3$ |
| Types of missingness for C1 in those with X=0 | Type 1: P(Miss)=0.3 |
| | Type 2: P(Miss)=$0.1 + 0.4*C2 + 0.4*C3$ |
| | Type 3: P(Miss)=$0.1 + 0.7*C1$ |
| | Type 4: P(Miss)=$0.3 + 0.2*Y$ |

**Table 2:** Risk difference estimates under each type of missingess in the untreated for the ATT and ATE across 2,000 replicates.

| Type of missingness | Mean ATT RD | ATT RD standard error | Mean ATE RD | Mean ATE RD standard error |
|---|---|---|---|---|
| No missingness | 0.220 | 0.0032 | 0.180 | 0.0029 |
| Type 1 missingness | 0.220 | 0.0036 | 0.185 | 0.0032 |
| Type 2 missingness | 0.220 | 0.0055 | 0.172 | 0.0036 |
| Type 3 missingness | 0.220 | 0.0049 | 0.174 | 0.0037 |
| Type 4 missingness | 0.271 | 0.0036 | 0.229 | 0.0031 |

ATT=average treatment effect in the treated; ATE=average treatment effect.

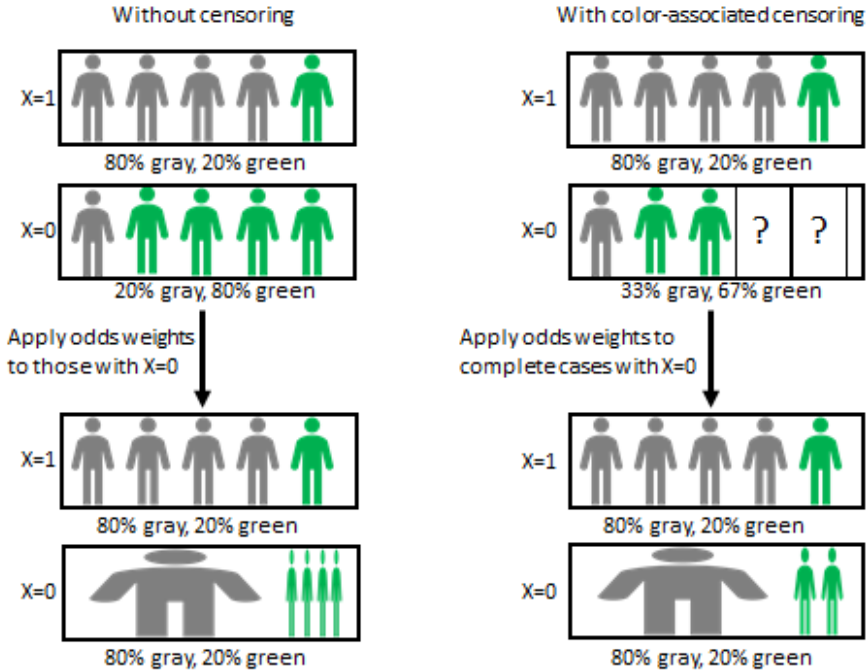

**Figure 2 Legend:** How CCA supplies an unbiased solution when those with X=0 have missing covariate data and the target population is those with X=1, demonstrated with odds weights. The final color distribution in the target of X=1 (80% gray and 20% green) remains constant. Even if the distribution of color in those with X=0 changes from 20% gray and 80% green to 33% gray and 67% green, with half of the green individuals selected out of the CCA due to missing data, the final distribution after weights is still 80% gray and 20% green. The only change is that the odds weight for green individuals changes from 1/4 to 1/2.

**Table 3:** Risk difference estimates under each type of missingess in the untreated for the AT-TRANS and AT-GEN across 2,000 replicates.

| Type of missingness | Mean AT-TRANS RD | AT-TRANS RD standard error | Mean AT-GEN RD | Mean AT-GEN RD standard error |
|---|---|---|---|---|
| No missingness | | | | |
| Type 1 missingness | | 0.0185 | 0.158 | 0.0181 |
| Type 2 missingness | | 0.0195 | 0.158 | 0.0192 |
| Type 3 missingness | | 0.0189 | 0.158 | 0.0185 |
| Type 4 missingness | | 0.0179 | 0.149 | 0.0176 |

AT-TRANS=average treatment effect in the external target population; AT-GEN=average treatment effect in the external target and study sample

**Table 4:** Implications of each type of covariate missingness in the unexposed/study population when estimating the ATT or AT-TRANS vs estimating the ATE or AT-GEN.

| Missingness type | Description | ATT/AT-TRANS | ATE/AT-GEN |
|---|---|---|---|
| Type 1 | Whether individuals are missing covariate data is completely random | No bias | Biased if effect differs between the treated and untreated |
| Type 2 | Whether individuals are missing covariate data depends on measured variables | No bias | Biased if the predictors of the missing covariate or treatment group are associated with treatment effect heterogeneity |
| Type 3 | Whether individuals are missing covariate data depends on the value of the variable itself | No bias | Biased if the missing covariate or treatment group are associated with treatment effect heterogeneity |
| Type 4 | Whether individuals are missing covariate data is independently* associated with the outcome | Biased | Biased |

ATT=average treatment effect in the treated; AT-EXT=average treatment effect in an external target; ATE=average treatment effect; AT-SUPER=average treatment effect in the superpopulation sampled for the study.
*Meaning an association after adjustment for measured covariates

Validity of Complete Case Analysis Depends on the Target Population

Supplemental Digital Content

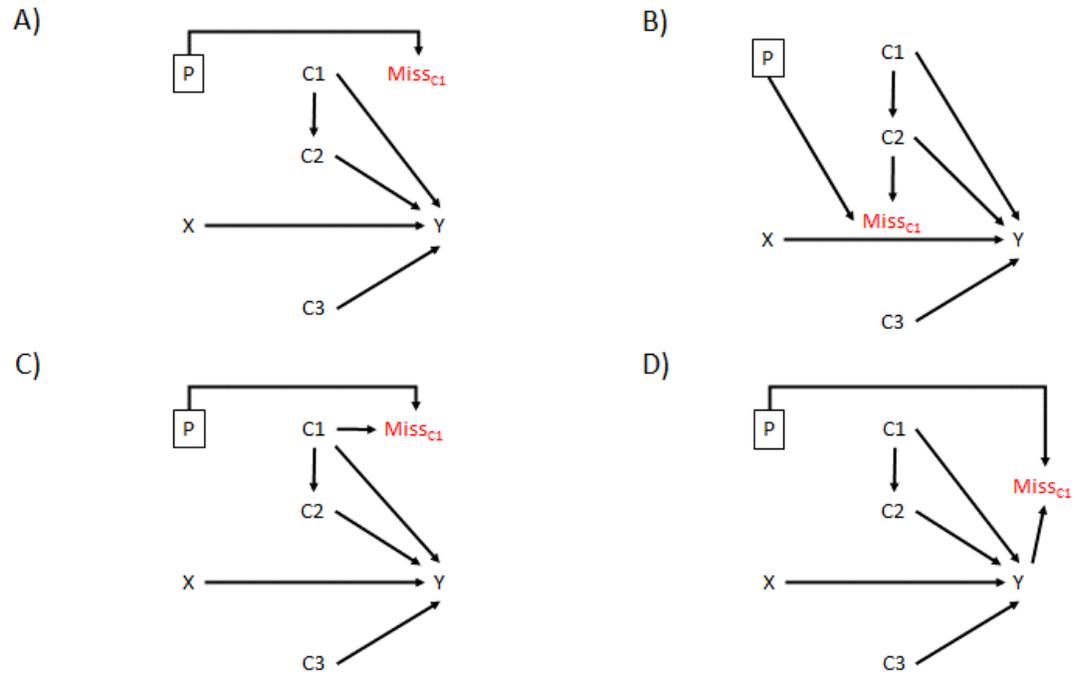

**Supplemental Figure S1 Legend:** Four directed acyclic graphs corresponding to four potential types of missingness of a covariate $M$ that is missing only within a trial population (represented by the P node). Type 1 corresponds to panel A, Type 2 to panel B, Type 3 to panel C, and Type 4 to panel D.

**Supplemental Table S1:** Simulation parameters for the simple ATT simulation

| Parameter | Value or value(s) |
|---|---|
| Number of individuals per simulation replicate | 10,000 |
| Number of replicates | 2,000 |
| P(C1) | 0.5 |
| P(C2 \| C1=1) | 0.3 |
| P(C2 \| C0=0) | 0.7 |
| P(C3) | 0.2 |
| P(S=1) | 0.2 + 0.15*C1 + 0.15*C2 + 0.15*C3 |
| P(X \| S=1) | 0.5 |
| P(Y=1) | 0.05 + 0.1*C1 + 0.1*C2 + 0.1*C3 + 0.1*C1*C2 + 0.05*X*C1 + 0.05*X*C2 + 0.05*X*C3 |
| Types of missingness for C1 in those with S=1 | Type 1: P(Miss)=0.3 |
| | Type 2: P(Miss)=0.3 + 0.1*C2 + 0.1*C3 |
| | Type 3: P(Miss)=0.3 + 0.1*C1 |
| | Type 4: P(Miss)=0.3 + 0.1*Y |

Estimand: ATT

Here transporting risk under no treatment to the treated

$P(Y^0 = 1 | X = 1)$

By law of total probability

$$= \sum_{z,m} P(Y^0 = 1 | X = 1, C = c, M = m) P(C = c, M = m | X = 1)$$

By conditional exchangeability $Y^a \coprod A | Z$

$$= \sum_{z,m} P(Y^0 = 1 | X = 0, C = c, M = m) P(C = c, M = m | X = 1)$$

By causal consistency

$$= \sum_{z,m} P(Y = 1 | X = 0, C = c, M = m) P(C = c, M = m | X = 1)$$

By $Y \coprod R_m | C, M, X = 0$ where $R = 1$ indicates a complete case

$$= \sum_{z,m} P(Y = 1 | X = 0, C = c, M = m, R_m = 1) P(C = c, M = m | X = 1)$$

By $C, M \coprod R | X = 1$

$$= \sum_{z,m} P(Y = 1 | X = 0, C = c, M = m, R_m = 1) P(C = c, M = m | X = 1, R_m = 1)$$